# Optical amplification and transmission of attenuated multi-soliton based on spectral characteristics of Akhmediev breather

Guangye Yang [a,b*], Fan O. Wu [b], Helena Lopez Aviles [b], and Demetrios Christodoulides [b]

[a] Department of Physics, Shanxi Medical University, Taiyuan, 030001, China
[b] CREOL, The College of Optics and Photonics, University of Central Florida, Orlando, Florida 32816, USA

*Corresponding author: ygy@sxmu.edu.cn

**Abstract:** We analyze temporal and spectral characteristics of Akhmediev breather and establish amplification and transmission of attenuated multi-soliton in nonlinear optical fiber. Our results show that the attenuated multi-soliton can be converted into Akhmediev breather through a judicious modulation of the spectrum. Subsequently, the maximally compressed pulse train of Akhmediev breather can be used to establish a robust breathing transmission by another spectrum modulation. In addition, the influence of the spectral modulation intensity on the excitation of Akhmediev breather and transmission of maximally compressed pulse train are also discussed.

## 1. Introduction

Akhmediev breather (AB) is a family of exact solutions of nonlinear Schrödinger (NLS) equation [1,2], which is periodic along the transverse dimension and is localized in longitudinal coordinate, and can be used to describe the evolution of initially continuous waves (CW) perturbed by a weak periodic modulation, i.e., modulation instability (MI) [3]. In such evolution, the CW with weak periodic perturbation will exponentially grow into a strongly compressed pulse train. However, such a pulse train cannot be preserved in the system due to the presence of a non-zero background. Thus, an interesting issue arises as to how to remove the non-zero background. So far, researchers are struggling to propose various feasible methods, such as polarization techniques [4], adiabatic amplification [5], Raman ffect [6], and delay line interferometer [7], etc.

Recently, based on AB solution, the generation and transmission of the high-repetition pulse train have been investigated via employing the delay-line interferometer approach [7,8]. Its dynamics in an optical fiber with a longitudinally tailored dispersion have been experimentally demonstrated, in which the breather evolution is nearly frozen at the maximum compression point [9]. The similar dynamics in dispersion exponentially decreasing fiber also has been investigated [10]. Also, the continuous wave supercontinuum generation from MI and Fermi-Pasta-Ulam (FPU) recurrence have been examined [11-13]. Besides, relevant theoretical and experimental works were reported in a quadratic medium and ferromagnet [14-17]. Another interesting issue is the limit of the AB solution. In this case, it can be reduced to a rational fraction solution for the NLS equation, i.e. the called Peregrine rogue solution. The Peregrine rogue wave is localized in both transverse and longitudinal dimensions and can be excited by the initial conditions of AB with a modulation frequency approaching one half [18]. As for applications, the Peregrine rogue wave can be used to generate a high-power pulse utilizing the spectral-filtering method [19,20].

In this paper, we study the potential applications of AB solution based on its temporal and spectral properties. Specifically, we focus on the amplification and transmission of an attenuated multi-soliton (AMS) in nonlinear optical fibers. Our results show that, by using a central spectral modulation method, the AMS can be extraordinarily amplified and robustly transmitted in realistic nonlinear dispersive optical fiber. We also discuss the influence of modulation intensity on the amplification and transmission processes. These results provide a better understanding of the underlying nonlinear mechanism of AB solution and the prospects of further improving the amplification techniques in optical fibers.

## 2. Spectral analysis of Akhmediev breather for NLS equation

The propagation of an optical pulse in a single-mode nonlinear optical fiber can be described by the following NLS equation [21]:

$$i\frac{\partial \psi}{\partial \xi} - \frac{\beta_2}{2}\frac{\partial^2 \psi}{\partial \tau^2} + \gamma |\psi|^2 \psi + i\frac{\alpha}{2}\psi = 0 \tag{1}$$

where $\psi(\xi,\tau)$ is the slowly varying amplitude of the pulse envelope, $\xi$ is the propagation distance and $\tau$ is the retarded time in a frame of reference moving with the group velocity $v_g$ ($\tau = t - \xi/v_g$). $\beta_2$ is the group velocity dispersion (GVD) coefficient, $\gamma$ is the Kerr nonlinear coefficient and $\alpha$ is the fiber loss. For the anomalous GVD and in the absence of the fiber loss, i.e., $\beta_2 < 0$ and $\alpha = 0$, Eq. (1) has AB solution of the form

$$\psi_{AB}(z,\tau) = \psi_{BW} + \psi_{MW} = \sqrt{P_0}\left(\phi_{BW} + \phi_{MW}\right)\exp(iz), \tag{2a}$$

where

$$\phi_{BW} = 1 - \frac{2(1-2a)\cosh(bz) + ib\sinh(bz)}{\sqrt{2a} + \cosh(bz)}, \tag{2b}$$

$$\phi_{MW} = \frac{\sqrt{2a}\left[2(1-2a)\cosh(bz) + ib\sinh(bz)\right]}{\sqrt{2a} + \cosh(bz)} \times \frac{1 + \cos(\omega\tau)}{\sqrt{2a}\cos(\omega\tau) - \cosh(bz)}. \tag{2c}$$

Here, $P_0$ is the averaged power, $z = (\xi - \xi_0)/L_{NL}$ is the normalized propagation distance with $\xi_0$ being an arbitrary constant and $L_{NL} = 1/\gamma P_0$ stands for the nonlinear length, $\omega = \sqrt{1-2a}\,\omega_c$ is the modulated frequency with $\omega_c = \sqrt{4\gamma P_0 / |\beta_2|}$, $a$ ($0 < a < 1/2$) is a real constant, and $b = \sqrt{8a(1-2a)}$ determines the instability growth rate.

Figure 1(a) shows the evolution of a AB solution, where realistic parameters of a standard silica SMF-28 fiber at 1550 nm have been used, i.e., $\beta_2 = -21.4$ ps$^2$/km and $\gamma = -1.2$ W$^{-1}$km$^{-1}$[22,23]. One can see that AB solution exhibits periodicity in the transverse dimension, meanwhile it is longitudinally localized. To further understand the AB solution, we decompose it into a superposition of a background wave (BW) and a modulation wave (MW), as shown in Eq. (2). Such decomposition is to remove the non-zero background from AB, in order to establish a robust transmission, and it has been previously derived from a Hirota equation in Refs. [24] and [25]. Here, the BW is only $\xi$-dependent, when $\xi < \xi_b = 5$ km it decreases with $\xi$ and when $\xi > \xi_b$ it increases, where distance $\xi_a$ is a reference point for better understanding the behavior of AB, as shown in Fig. 1(b). The MW and it is a temporal periodic function with a period of $T_{AB} = 2\pi/\omega$ and has no background, as shown in Fig. 1(c). One should in mind that, comparing the AB and MW, the peak power of MW is bigger than the AB at maximally compressed point $\xi_b = 5$ km. This results from the superposition of MW and BW at this point.

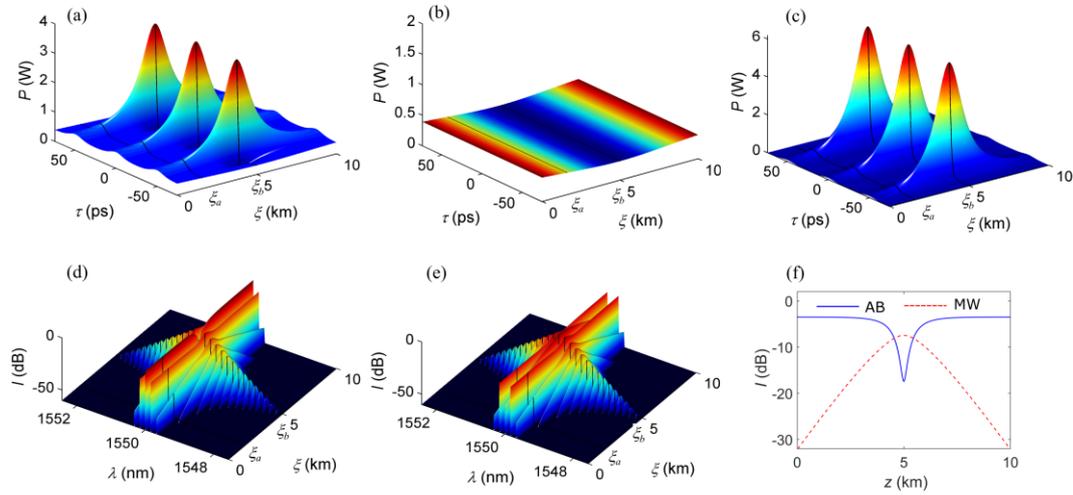

Fig. 1 Temporal evolution of (a) $\psi_{AB}$; (b) $\psi_{BW}$, and (c) $\psi_{MW}$ given by Eq. (2), corresponding the spectral evolution of AB (d) and MW (e),. (f) Evolution of central spectral intensity of AB ($\tilde{A}_{0AB}$) and MW ($\tilde{A}_{0MW}$) given by Eq. (3a) and (3b), respectively. The parameters used here are $P_0 = 0.45$ W, $\xi_0 = 5$, $a = 0.42$, $T_{AB} = 49.4$ ps, $\xi_a = 1.62$ km and $\xi_b = 5$ km.

Fig. 1(d) and 1(e) show the spectral evolution of AB and MW, respectively. The evolution of AB and MW, however similar, they are fundamentally different when careful analysis is made. In other words, from the Fourier expansions of AB and MW given by Eq.(2), we can have a clear

look at their pump and harmonic amplitudes:

$$\tilde{A}_{0AB}(z) = 1 - \frac{2(1-2a)\cosh(bz) + ib\sinh(bz)}{\sqrt{\cosh^2(bz) - 2a}}, \quad (3a)$$

$$\tilde{A}_{0MW}(z) = \tilde{A}_{0AB}(z) - \phi_{BW}(z), \quad (3b)$$

$$\tilde{A}_{nAB}(z) = \tilde{A}_{nMW}(z) = \left[1 - \tilde{A}_{0AB}(z)\right] \times \left[\frac{\cosh(bz) - \sqrt{\cosh^2(bz) - 2a}}{\sqrt{2a}}\right]^{|n|}. \quad (3c)$$

where $\tilde{A}_{0AB}$ and $\tilde{A}_{0MW}$ denote the pump amplitudes for AB and MW, respectively. $\tilde{A}_{nAB}$ and $\tilde{A}_{nMW}$ ($n = \pm 1, \pm 2, \cdots$) are the spectral harmonic amplitudes of the $n$th sideband. Note that the factors of the constant amplitude and phase in Eq. (3) are ignored [26,27]. From Eq. (3c), one finds that, the amplitudes of the $n$th sidebands for AB and MW are the same, but their pump amplitudes are different. In this case, the spectra of AB and MW are different only at the central frequency, as shown in Fig. 1(f). Therefore, one can transform AB into MW (or going back from MW to AB) by judiciously adjusting their pump at the central frequency.

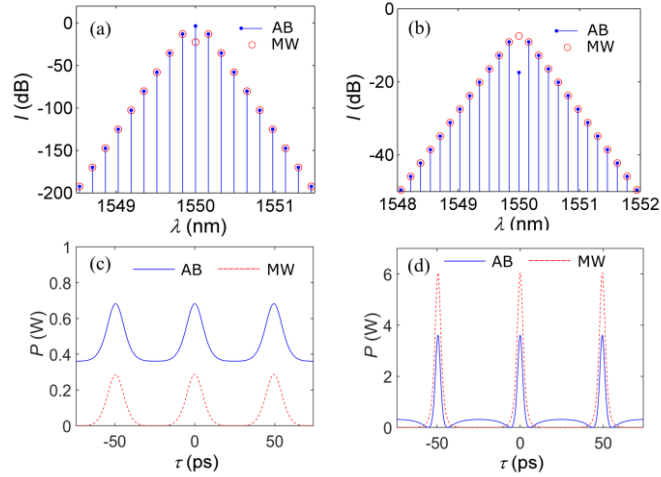

Fig. 2 The spectral intensity profiles of AB and MW at (a) $\xi_a = 1.62$ km and (b) $\xi_b = 5$ km, (c) and (d) are the corresponding power profiles, respectively.

In order to clarify this mechanism, Fig. 2 presents the profiles of the spectral intensity and temporal profiles of AB and MW at distances $\xi_a = 1.62$ km and $\xi_b = 5$ km, respectively. As indicated by Eq. (3), the spectral distributions of AB and MW are different only at the central frequency, as shown in Figs. 2(a) and 2(b). Thus, by modulating the pump of the MW train at the central frequency, it can be transformed into AB train [see Fig. 2(c)]. In the same vein, AB train can be transformed back to the MW train [see Fig. 2(d)]. Note that AB is characterized by its spatial localization property due to the underlying nonlinear interactions between many frequency components. However, in this case, the destructive evolution of AB is hindered since the modulated central frequency in MW redirects the coherent nonlinear couplings between the frequency components. Based on the spectral characteristics, one can achieve an amplification and propagation of pulse train by two successive pump modulation. The first modulation leads to a transformation of MW into AB, forming a high-power compressed pulse train with non-zero background due to the characteristic behavior of AB. The second modulation transforms the peaks of AB into MW. Thus, the pulse train eliminates the non-zero background and can avoid FPU recurrence, establishing a robust transmission.

### 3. Amplification and transmission of an attenuated multi-soliton

To demonstrate the amplification and transmission, we consider a multi-soliton (MS) as an initial input

$$\psi(0,\tau) = \sum_{j=-n}^{n} \sqrt{P_\mathrm{I}} \, \mathrm{sech}\left(\frac{\tau \pm jT_\mathrm{MS}}{T_0}\right), \tag{4}$$

where $P_\mathrm{I}$ is initial peak power. $T_0 = \left[|\beta_2|/(\gamma P_0)\right]^{1/2}$ is the temporal scale. $n$ is a positive integer number and $T_\mathrm{MS}$ represents the separation between neighboring solitons. In the presence of loss (here we take $\alpha = 0.19$ dB/km [22,23]), during propagation, the power of each pulse is decreased exponentially, meanwhile, the width is broadened. After a certain distance, the initial input evolved into a low power pulse train, i.e., AMS. In this case, we need to amplify the low power train in order to realize a robust transmission.

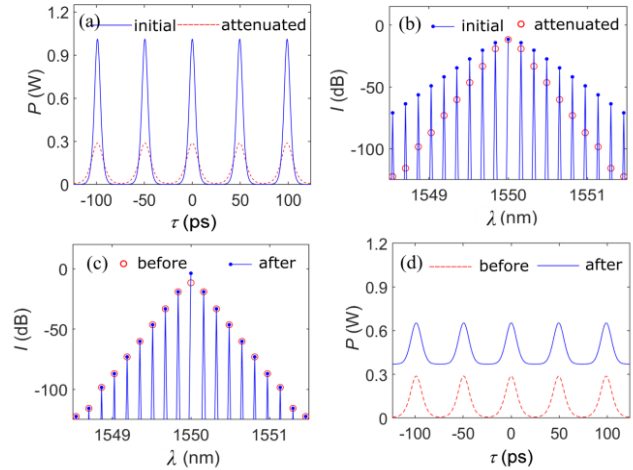

Fig. 3 (a) Temporal profiles of the initial MS given by Eq.(4) and the AMS after propagating $\xi_\mathrm{L} = 13.56$ km, and (b) corresponding spectral profiles. (c) Spectral profiles of the AMS before and after central spectral modulation with $G_\mathrm{M1} = 7.9$ dB, and (d) corresponding temporal profiles. Here, the parameters are $P_\mathrm{I} = 1$ W, $T_\mathrm{MS} = T_\mathrm{AB}$, and $n = 2$.

Figures. 3(a) and 3(b) show the power distribution of the initial MS given by Eq. (4) and the AMS in the temporal and spectral domains, respectively. In this case, the peak power is decreased from 1W to 0.28W after $\xi_\mathrm{L} = 13.56$ km of propagation. Interestingly, even though the intensity in each sideband is reduced as a result of attenuation, the strength of the pump remained unaffected. We will show that, the AMS can be amplified by only increasing the pump intensity with $G_{M1} = 7.9\text{dB}$ as shown in Fig. 3(c). Thus, the minimum power of the AMS $P_{BG1}$ is enhanced from $0\text{W}$ to $0.37\text{W}$, as shown in Fig. 3(d). As a result, the modified AMS can be treated as a superposition of a periodic modulation wave and a continuous wave with $0.37\text{W}$, and hence it matches the excitation condition of AB. Therefore, the AMS, which is now sitting on the CW background undergoes a strong temporal compression. As a result, the peak-power increases until the maximum compression of the pulse train is reached at a distance $\xi_P = 3.74\text{km}$. Such a maximally compressed pulse train (MCPT) has a non-zero background, so that it will eventually disappear due to the nature of AB, as shown in Fig. 4(a). Figure 4(b) presents the evolution of the corresponding spectrum, which starts with narrow spectral components and then spreads into a triangular-type shape, then shrinks to recover the initial shape. In order to attain a robust transmission of the soliton train in optical fibers, one must eliminate the non-zero background of the MCPT. According to the above spectral characteristics of AB, we can eliminate the background wave by appropriately modify the pump component of the field with a $\pi$ phase shift and an intensity amplification of $G_{M2} = 10\text{dB}$. Figs. 4(c) and 4(d) present the spectrum and the corresponding temporal power profile of the MCPT before and after the modulation, respectively. One can see that the intensity at the central frequency is enhanced and the background wave is eliminated. The MCPT becomes a pulse train with a higher power, which is by itself holding a similar form of the MW, as shown in Fig. 2(b) and 2(d).

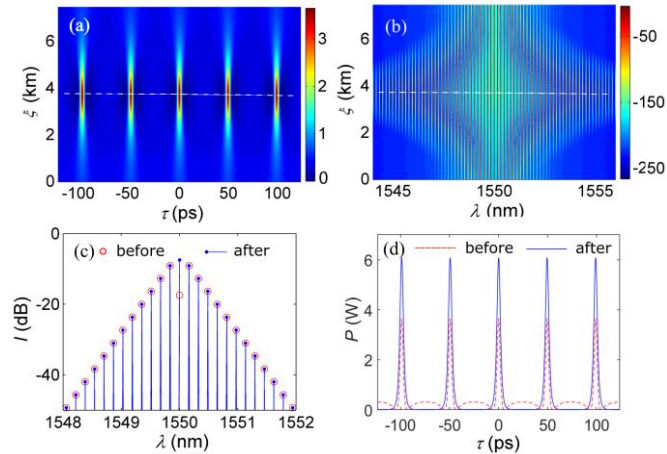

Fig. 4 The evolution of (a) the AB excited by the AMS after spectral modulation, and (b) corresponding spectral evolution. (c) Spectral Profiles of the MCPT before and after central spectral modulation with $G_{M2} = 10\text{dB}$ at maximally compressed distance $\xi_P = 3.74\text{km}$ (white dotted lines in (a) and (b)), and (d) corresponding temporal profiles. Parameters of the initial condition are the same as Fig.3.

Figure. 5(a) demonstrates the power evolution in both time and frequency domains during the whole amplification process. In such a process, two subsequent central frequency modulation were applied: the first one at $\xi = 0\text{km}$ was intended to convert AMS into AB; the second one at

$\xi_P = 3.74\text{km}$ was to prevent AB from evolving back to CW. As a result, the AMS was preserved and a stable propagation was re-established, each pulse exhibits the behavior of a breathing soliton. Moreover, the stable breathing behavior is further justified by the corresponding spectral evolution, as shown in Fig.5(b), where a comb-like intensity distribution is evolving in a clear periodic manner-free from any destructive dynamics like soliton fission, etc.

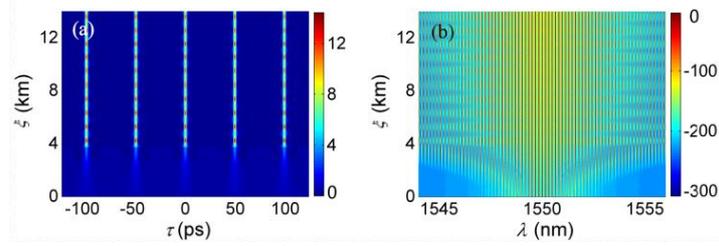

Fig. 5 (a) The evolution of the AMS that experiences two spectral modulations with $G_{M1} = 7.9\text{dB}$ and $G_{M2} = 10\text{dB}$ at $\xi = 0\text{km}$ and $\xi_P = 3.74\text{km}$, respectively. (b) Corresponding spectral evolution which shows a typical breathing characteristic. Initial conditions are the same as Fig.3.

## 4. Influence of spectral modulation intensities on excitation of AB and amplification of the AMS

While the scheme demonstrated above can be used to conveniently retrieve AMS by applying two subsequent central spectral modulations, the strength of spectral modulation intensity $G_{M1}$ and $G_{M2}$ play crucial roles in such processes. Here, we discuss the impact of $G_{M1}$ and $G_{M2}$ on the MCPT of AB and the final amplification of AMS.

Figures. 6(a) and 6(b) show the temporal power profiles of the AMS after the first modulation and the corresponding MCPT with different $G_{M1}$. As $G_{M1}$ increases, the background of the AMS and the peaks of the MCPT become stronger, meanwhile, the compression is accelerated. This is due to the fact that the coupling coefficients of nonlinear four-wave mixing between pump and sidebands are proportional to the pump intensity, hence, the background power of the AMS $P_{BG1}$ is inversely proportional to the position of the maximum compression distance $\xi_P$, as shown in Fig.6(c). Interestingly, from Fig. 6(d), one can see that when $G_{M1}$ increases, the peak power of the MCPT $P_{P1}$ grows faster than the background power of the MCPT $P_{BG2}$. It indicates that most of the extra power will be transferred to the pulses, not the background, thus enable us to attain high efficiency transmission.

Figures. 6(e) and 6(f) depict the temporal power profiles when $\xi_P = 3.74\text{km}$ and the evolution of the peak power along with propagation distance, after the second spectral modulation with different $G_{M2}$. More importantly, with increasing $G_{M2}$, the peak power $P_{P2}$ and width $T_W$ of the MCPT after spectral modulation also increase, associated with a faster oscillation, as shown in Figs. 6(g) and 6(h). In addition, one can see that the background power $P_{BG3}$ has no significant change. Therefore, by adjusting the modulation strength, the amplification scheme of AMS provided in this paper has a broad range of tunability in terms of pulse width and peak power.

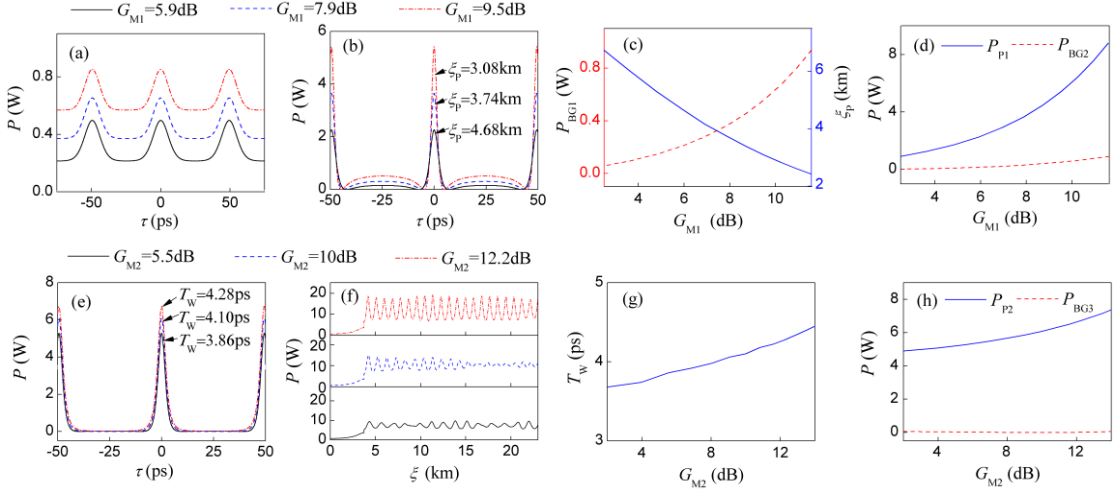

Fig. 6 (a,b)Spectral Profiles of the AMS after first spectral modulation with different $G_{M1}$ and corresponding MCPT.(c,d) Dependence of the background power of the AMS $P_{BG1}$, the position of MCPT $\xi_P$, and the peak power $P_{P1}$ and background power $P_{BG2}$ of MCPT on the first spectral modulation intensity $G_{M1}$. (e,f)Spectral profiles of MCPT ($G_{M1} = 7.9\text{dB}$) after second spectral modulation with different $G_{M2}$ and evolutions of corresponding peak power. (g,h)Dependence of the width $T_W$, peak power $P_{P2}$ and background power $P_{BG3}$ of MCPT on the second spectral modulation intensity $G_{M2}$. Here, the parameters of the initial condition are the same as Fig.3.

## 5. Conclusions

In summary, we have studied the time and frequency domain characteristics of Akhmediev breather and have established amplification and transmission of an attenuated multi-soliton by applying two subsequent central spectral modulations in optical fibers. In such a scheme, first the attenuated multi-soliton is charged by the modulation background due to the build-up process of Akhmediev breather and then the background is modulated again right at the maximum compression point in order to prevent the upcoming collapsing process of such localized breathers. The output pulses exhibit a robust breathing propagation feature. Finally, the impacts of different levels of modulation strength of the two spectrum modulations have also been investigated. Our results provide constructive guidance for the application of various optical breathers (Akhmediev breather, superregular breathers [28,29], doubly periodic breathers [30], etc.) and could be implemented to preserve high-repetition pulse train in optical fibers.

**Acknowledgements**

This research is supported by the National Natural Science Foundation of China grant (No.61505101), the Natural Science Foundation of Shanxi province (No.201901D111212), the Doctoral Startup Research Fund of Shanxi Medical University (No.03201401), the Office of Naval Research of USA (No. MURI N00014-17-1-2588) and National Science Foundation of USA (No.ECCS-1711230).